\documentstyle[aps]{revtex}

\draft

\begin{document}

\tighten
\twocolumn[\hsize\textwidth\columnwidth\hsize\csname@twocolumnfalse\endcsname

\title{Coexisting Kondo singlet state with antiferromagnetic long-range
order: A possible ground state for Kondo insulators}
\author{Guang-Ming Zhang$^1$ and Lu Yu$^{2,3}$}
\address{$^1$ Center for Advanced Study, Tsinghua University, Beijing 100084,
P. R. China\\
$^2$ International Center for Theoretical Physics, P. O. Box 586, Trieste
34100, Italy\\
$^3$ Institute for Theoretical Physics, Academic Sinica, Beijing 100080,
P. R. China}
\date{\today}
\maketitle

\begin{abstract}
The ground-state phase diagram of a half-filled anisotropic Kondo lattice
model is calculated within a mean-field theory. For small transverse exchange
coupling $J_{\perp }<J_{\perp c1}$, the ground state shows an
antiferromagnetic long-range order with finite staggered
magnetizations of both localized spins and conduction electrons. When
$J_{\perp }>J_{\perp c2}$, the long-range order is destroyed  and
the system is in a disordered Kondo singlet state with a hybridization
gap. Both ground states can describe the  low-temperature
phases of Kondo insulating compounds. Between these two distinct phases,
 there may be a coexistent regime as a result of the balance
between local Kondo screening and magnetic interactions.
\newline
{PACS numberes: 71.28.+d, 72.15.Qm, 75.20.Hr}\newline
\end{abstract}

]

The Kondo lattice model is often considered as a theoretical model for heavy
fermion materials. For this model, an important issue arises from the
interplay between the Kondo screening and the magnetic interactions among
localized spins mediated by the conduction electrons. The former effect
favors a nonmagnetic Kondo singlet state in strong coupling limit, while the
latter interactions tend to stabilize a magnetically long-range ordered
state in weak coupling limit. The nature of such a transition between these
two distinct phases has been a long standing issue since it was first
suggested by Doniach \cite{doniach}. The one-dimensional model was
intensively studied at half filling, showing that its ground state is always
a disordered Kondo singlet state (or a spin liquid state)\cite{review}. In
higher dimensions, however, both antiferromagnetic long-range order (AFLRO)
and disordered Kondo singlet states may occur \cite{lacy}. Recently, there
have been more indications of such a transition from various approximate
treatments for the Kondo or Anderson lattice models, including variational
Monte Carlo calculation \cite{wang}, higher-order series expansions \cite
{shi}, quantum Monte Carlo simulations \cite{vekic,assaad}, and infinite
dimensional calculations \cite{jarrell}.

In this paper, we would like to consider the issue whether the disordered
Kondo singlet state can coexist with an AFLRO at half-filling. First of all,
we introduce an anisotropic Kondo lattice model by distinguishing the
longitudinal spin exchange interaction from the transverse one, because we
notice that the longitudinal interaction describes a polarization of the
conduction electrons by the localized spins, while the transverse one
describes a spin-flip scattering of the conduction electrons off the
localized spins. The former interaction is the origin of the magnetic
interactions among localized spins, leading to an AFLRO; the latter
interaction is responsible for the local Kondo screening effects, yielding a
disordered Kondo singlet state. We also notice that the AFLRO and local
Kondo singlet order operators form an irreducible representation of an SU(2)
algebra, which can be regarded as the spectrum generating algebra \cite{sga}
of the half-filled Kondo lattice model. Within the framework of a mean field
theory, the magnetic interactions and the Kondo screening are considered on
an equal footing, and the ground-state phase diagram of the model
Hamiltonian is calculated.

The symmetric Kondo lattice model with anisotropic exchange couplings is
defined as: 
\begin{eqnarray}
H &=&\sum_{{\bf k}\sigma }\epsilon _{{\bf k}}c_{{\bf k}\sigma }^{\dagger }c_{%
{\bf k}\sigma }+H_{\parallel }+H_{\perp }  \nonumber \\
H_{\parallel } &=&\frac{J_{\parallel }}4\sum_i(d_{i\uparrow }^{\dagger
}d_{i\uparrow }-d_{i\downarrow }^{\dagger }d_{i\downarrow })(c_{i\uparrow
}^{\dagger }c_{i\uparrow }-c_{i\downarrow }^{\dagger }c_{i\downarrow }) 
\nonumber \\
H_{\perp } &=&\frac{J_{\perp }}2\sum_i(d_{i\downarrow }^{\dagger
}d_{i\uparrow }c_{i\uparrow }^{\dagger }c_{i\downarrow }+d_{i\uparrow
}^{\dagger }d_{i\downarrow }c_{i\downarrow }^{\dagger }c_{i\uparrow }),
\end{eqnarray}
where the pseudo-fermion representation for the localized spins $%
S_i^z=(d_{i\uparrow }^{\dagger }d_{i\uparrow }-d_{i\downarrow }^{\dagger
}d_{i\downarrow })/2,$ $S_i^{-}=d_{i\downarrow }^{\dagger }d_{i\uparrow },$ $%
S_i^{+}=d_{i\uparrow }^{\dagger }d_{i\downarrow },$ has been used with a
local constraint $d_{i\uparrow }^{\dagger }d_{i\uparrow }+d_{i\downarrow
}^{\dagger }d_{i\downarrow }=1$. Note that $H_{\perp }$ can also be
rewritten in the form of 
\[
\frac{-J_{\perp }}4\sum_i\left[ (d_{i\downarrow }^{\dagger }c_{i\downarrow
}+c_{i\uparrow }^{\dagger }d_{i\uparrow })^2+(d_{i\uparrow }^{\dagger
}c_{i\uparrow }+c_{i\downarrow }^{\dagger }d_{i\downarrow })^2\right] . 
\]
$H_{\parallel }$ describes the polarization of conduction electrons by the
local impurity spins, leading to a magnetic instability, while $H_{\perp }$
corresponds to the spin-flip scatterings, giving rise to the local Kondo
screening effect. The latter effect has been investigated by various
approaches, in particular, those based on a 1/N expansion \cite{1/N} (N is
the degeneracy of the localized spin). However, the former effect is not
treated on an equal footing in these approaches because the magnetic
interaction occurs there only at 1/N$^2$ order \cite{1/N2}.

It has been known that the spectrum generating algebra plays an important
role in analyzing complete spectra for a model. For instance, it has been
used to study collective excitations and phase transitions in
one-dimensional metals and two-dimensional Hubbard model \cite{sga}. Here
the spectrum generating algebra for the half-filled Kondo lattice model is
given by generators of an SU(2) Lie group: 
\begin{eqnarray}
\tau ^{+} &=&\sum_ie^{i{\bf Q}\cdot {\bf r}_i}(c_{i\uparrow }^{\dagger
}d_{i\uparrow }+d_{i\downarrow }^{\dagger }c_{i\downarrow }),  \nonumber \\
\tau ^{-} &=&\sum_ie^{-i{\bf Q}\cdot {\bf r}_i}(d_{i\uparrow }^{\dagger
}c_{i\uparrow }+c_{i\downarrow }^{\dagger }d_{i\downarrow }),  \nonumber \\
\tau ^z &=&\frac 12\sum_i\left[ (c_{i\uparrow }^{\dagger }c_{i\uparrow
}-c_{i\downarrow }^{\dagger }c_{i\downarrow })-(d_{i\uparrow }^{\dagger
}d_{i\uparrow }-d_{i\downarrow }^{\dagger }d_{i\downarrow })\right] ,
\end{eqnarray}
which satisfy the commutation relations: $\left[ \tau ^z,\tau ^{\pm }\right]
=\pm \tau ^{\pm },$ $\left[ \tau ^{+},\tau ^{-}\right] =2\tau ^z,$ the SU(2)
algebra. ${\bf Q}$ is the AF reciprocal vector. We also find that an
irreducible representation of this SU(2) algebra can serve as the order
parameter operators for this model. They are given by: 
\begin{eqnarray}
K^{+} &=&\sum_i(c_{i\uparrow }^{\dagger }d_{i\uparrow }+d_{i\downarrow
}^{\dagger }c_{i\downarrow }),  \nonumber \\
K^{-} &=&-\sum_i(d_{i\uparrow }^{\dagger }c_{i\uparrow }+c_{i\downarrow
}^{\dagger }d_{i\downarrow }),  \nonumber \\
M^z &=&\frac 12\sum_ie^{i{\bf Q}\cdot {\bf r}_i}\left[ (d_{i\uparrow
}^{\dagger }d_{i\uparrow }-d_{i\downarrow }^{\dagger }d_{i\downarrow
})-(c_{i\uparrow }^{\dagger }c_{i\uparrow }-c_{i\downarrow }^{\dagger
}c_{i\downarrow })\right] ,  \nonumber
\end{eqnarray}
with the following commutation relations $\left[ \tau ^{\pm },K^{\mp
}\right] =2M^z,$ $\left[ \tau ^{\pm },M^z\right] =K^{\pm }$, $\left[ \tau
^z,K^{\pm }\right] =\pm K^{\pm },$ and $\left[ \tau ^z,M^z\right] =0,$ where 
$K^{\pm }$ are the order parameter operators describing the local Kondo
singlet state and $M^z$ is the staggered magnetization operator describing
the AFLRO in terms of a commensurate spin-density wave. The expectation
values of these order operators are the corresponding order parameters. Note
that {\it the model Hamiltonian does not satisfy this SU(2) symmetry defined
by the spectrum generating algebra,} which can however provide us with a
guide to choose the order parameters in the model Hamiltonian.

Now we introduce a mean-field theory treating equally the magnetic
interactions between the localized spins and the itinerant electron
screening aspects. The longitudinal interaction term is approximated as: 
\begin{eqnarray*}
&&\frac{J_{\parallel }}2\left[ m_d\sum_ie^{i{\bf Q}\cdot {\bf r}%
_i}(c_{i\uparrow }^{\dagger }c_{i\uparrow }-c_{i\downarrow }^{\dagger
}c_{i\downarrow })\right. \\
&&\text{ \qquad }\left. -m_c\sum_ie^{i{\bf Q}\cdot {\bf r}_i}(d_{i\uparrow
}^{\dagger }d_{i\uparrow }-d_{i\downarrow }^{\dagger }d_{i\downarrow
})\right] +J_{\parallel }m_cm_d{\cal N},
\end{eqnarray*}
where the staggered magnetizations have been introduced as the AF order
parameters $m_c=-\frac 12<c_{i\uparrow }^{\dagger }c_{i\uparrow
}-c_{i\downarrow }^{\dagger }c_{i\downarrow }>e^{-i{\bf Q}\cdot {\bf r}_i}$
and $m_d=\frac 12<d_{i\uparrow }^{\dagger }d_{i\uparrow }-d_{i\downarrow
}^{\dagger }d_{i\downarrow }>e^{-i{\bf Q}\cdot {\bf r}_i},$ and a minus sign
has been absorbed into the definition of $m_c$ for convenience. The
transverse interaction term is approximated to be: 
\[
\frac{J_{\bot }V}2\sum_{i\sigma }(c_{i\sigma }^{\dagger }d_{i\sigma
}+d_{i\sigma }^{\dagger }c_{i\sigma })+\frac{J_{\bot }}2V^2{\cal N}, 
\]
with the hybridization order parameter describing the local Kondo singlet
state $-V=<c_{i\uparrow }^{\dagger }d_{i\uparrow }+d_{i\downarrow }^{\dagger
}c_{i\downarrow }>$ $=<d_{i\uparrow }^{\dagger }c_{i\uparrow
}+c_{i\downarrow }^{\dagger }d_{i\downarrow }>.$ Moreover, we could also
introduce a $d$-electron chemical potential term $\Sigma _{i,\sigma }$ $%
E_dd_{i\sigma }^{\dagger }d_{i\sigma }$ to fix the $d$-electron density to
be one on each site, but actually this is not necessary at half filling
because the electron-hole symmetry automatically imposes $E_d=0$ \cite
{review}. Therefore, the mean field Hamiltonian can be written in the
following form 
\begin{equation}
H=\sum_{{\bf k}\sigma }{}^{^{\prime }}\Psi _{{\bf k}\sigma }^{\dagger }%
\widehat{H}_{{\bf k}\sigma }\Psi _{{\bf k}\sigma }+{\cal N}(J_{\parallel
}m_cm_d+\frac{J_{\bot }V^2}2),
\end{equation}
where $\Psi _{{\bf k}\sigma }^{\dagger }=(c_{{\bf k}\sigma }^{\dagger },$ $%
c_{{\bf k}+{\bf Q}\sigma }^{\dagger },$ $d_{{\bf k}\sigma }^{\dagger },$ $d_{%
{\bf k}+{\bf Q}\sigma }^{\dagger })$, its transposition $\Psi _{{\bf k}%
\sigma }$, and the matrix 
\begin{equation}
\widehat{H}_{{\bf k}\sigma }=\left[ 
\begin{array}{llll}
\epsilon _{{\bf k}}, & \frac{J_{\parallel }}2m_d\sigma , & \frac{J_{\bot }V}2%
, & 0 \\ 
\frac{J_{\parallel }}2m_d\sigma , & -\epsilon _{{\bf k}}, & 0, & \frac{%
J_{\bot }V}2 \\ 
\frac{J_{\bot }V}2, & 0, & 0, & -\frac{J_{\parallel }}2m_c\sigma \\ 
0, & \frac{J_{\bot }V}2, & -\frac{J_{\parallel }}2m_c\sigma , & 0
\end{array}
\right] .
\end{equation}
The static staggered magnetizations partially break the translational
symmetry and the Brillouin zone is folded in half, so that the summation
over ${\bf k}$ is taken in the reduced Brillouin zone. The quasiparticle
bands are determined by the equation: $\mid E\widehat{I}-\widehat{H}\mid =0,$
giving rise to four bands with dispersions, 
\[
E_{\pm \pm }({\bf k})=\pm \frac 1{\sqrt{2}}\sqrt{\epsilon _{{\bf k}%
}^2+J_{\parallel }^2(m_c^2+m_d^2)/4+J_{\bot }^2V^2/2\pm E^{\prime }({\bf k})}
\]
with 
\begin{eqnarray*}
E^{\prime }({\bf k}) &=&\left\{ \left[ \epsilon _{{\bf k}}^2+J_{\parallel
}^2(m_c^2+m_d^2)/4+J_{\bot }^2V^2/2\right] ^2\right. \\
&&\left. -\frac 14\left( J_{\parallel }^2m_cm_d+J_{\bot }^2V^2\right)
^2-J_{\parallel }^2m_c^2\epsilon _{{\bf k}}^2\right\} ^{1/2},
\end{eqnarray*}
and the ground-state energy is given by 
\[
E_g=2\sum_{{\bf k}}{}^{^{\prime }}\left[ E_{--}({\bf k})+E_{-+}({\bf k}%
)\right] +{\cal N}(J_{\parallel }m_cm_d+\frac{J_{\bot }V^2}2), 
\]
corresponding to completely filling the two negative energy bands with
electrons. The self-consistent equations are obtained by minimizing the
ground-state energy $E_g$ with respect to $m_d$, $m_c$, and $V$,
respectively. After the summations over momenta are transformed into
integrals over energies by assuming a constant density of states of the
conduction electrons in $[-D,D]$, the self-consistent equations are
expressed in the form 
\begin{eqnarray}
&&\frac{\lambda _{\parallel }}8%
\displaystyle \int %
\limits_{-1}^1d\varepsilon \frac{\text{ }2F_2(\varepsilon )m_d+(\lambda
_{\parallel }^2m_cm_d+\lambda _{\bot }^2V^2)m_c}{2F_1(\varepsilon
)F_2(\varepsilon )}=m_c,  \nonumber \\
&&\frac{\lambda _{\parallel }}8%
\displaystyle \int %
\limits_{-1}^1d\varepsilon \frac{\text{ }2(F_2(\varepsilon )+2\varepsilon
^2)m_c+(\lambda _{\parallel }^2m_cm_d+\lambda _{\bot }^2V^2)m_d}{%
2F_1(\varepsilon )F_2(\varepsilon )}=m_d,  \nonumber \\
&&\frac{\lambda _{\perp }}4%
\displaystyle \int %
\limits_{-1}^1d\varepsilon \frac{2F_2(\varepsilon )+(\lambda _{\parallel
}^2m_cm_d+\lambda _{\bot }^2V^2)}{2F_1(\varepsilon )F_2(\varepsilon )}=1,
\end{eqnarray}
where we have used the simplified expressions in terms of $F_1$ and $F_2,$ 
\begin{eqnarray}
F_1(\varepsilon ) &=&\sqrt{\varepsilon ^2+\lambda _{\parallel
}^2(m_c^2+m_d^2)/4+\lambda _{\bot }^2V^2/2+F_2(\varepsilon ),}\newline
\nonumber \\
F_2(\varepsilon ) &=&\sqrt{\frac 14\left( \lambda _{\parallel
}^2m_cm_d+\lambda _{\bot }^2V^2\right) ^2+\lambda _{\parallel
}^2m_c^2\varepsilon ^2},
\end{eqnarray}
and we have also used the notations: $\lambda _{\parallel }=J_{\parallel
}/D, $ $\lambda _{\perp }=J_{\perp }/D$, where $2D$ is the bandwidth of the
conduction electrons.

When $\lambda _{\parallel }\gg \lambda _{\perp }$, the polarization effect
described by $H_{\parallel }$ dominates, and the ground state has an AFLRO
characterized by finite staggered magnetizations for both localized spins
and the conduction electrons. Thus, the total staggered magnetization of the
system should be given by $M^z/{\cal N}=(m_d+m_c)$, while the staggered
magnetic moment is $\mu =(m_d-m_c)$. Let us assume for the moment they are
the only order parameters, {\it i.e.} $V=0$. All equations involved are
greatly simplified. Due to unit cell doubling the conduction electron band
is folded in half to form two new bands: $\epsilon _{\pm }({\bf k})=\pm 
\sqrt{\epsilon _{{\bf k}}^2+(J_{\parallel }m_d)^2/4}.$ There is a gap in the
conduction electron excitation spectrum: $\Delta =J_{\parallel }m_d/2.$ The
charge gap is twice this value, $\Delta _{ch}=J_{\parallel }m_d$. Moreover,
the ground-state energy of the AFLRO state can also be found 
\begin{equation}
\frac{E_g^{AF}}{{\cal N}D}=-\frac 12\left[ \sqrt{1+\left( \frac{\lambda
_{\parallel }}4\right) ^2}+\left( \frac{\lambda _{\parallel }}4\right) \frac{%
m_c}{m_d}\right] ,
\end{equation}
where ${\cal N}$ is the total number of the lattice sites, and $\frac{m_c}{%
m_d}=\frac{\lambda _{\parallel }}4\ln \left( \frac 4{\lambda _{\parallel }}+%
\sqrt{1+(\frac 4{\lambda _{\parallel }})^2}\right) $with $m_d=\frac 12$.

In the opposite limit, when $\lambda _{\perp }\gg \lambda _{\parallel }$,
the Kondo screening effect described by $H_{\perp }$ dominates, and the
ground state has a finite local Kondo order parameter $V$. If we assume
again for the moment it is the only order parameter, the system is a
nonmagnetic band insulator with an effective hybridization: $\frac 1V%
=\lambda _{\perp }\sinh \left( \frac 1{\lambda _{\perp }}\right) $. A
similar result has already been given by the 1/N expansion approach \cite
{1/N}, which is believed to describe the correct low-energy Kondo physics.
The quasiparticle excitation spectrum is expressed in the form $\epsilon
_{\pm }^{\prime }({\bf k})=\frac 12\left( \epsilon _{{\bf k}}\pm \sqrt{%
\epsilon _{{\bf k}}^2+\left( J_{\perp }V\right) ^2}\right) ,$ and there is a
small hybridization gap $\Delta _{hy}\simeq \left( J_{\perp }V\right) ^2/2D$%
, which splits the Kondo resonance formed at the Fermi level. At half
filling, each lattice site has one conduction electron and one localized
spin, and they can form a local singlet state, thus the system becomes a
disordered nonmagnetic insulator, a band-insulator with both charge and spin
gap $2\Delta _{hy}$. The ground-state energy can be also calculated 
\begin{equation}
\frac{E_g^{KS}}{{\cal N}D}=-\frac 12\coth \left( \frac 1{\lambda _{\perp }}%
\right) .
\end{equation}
Physically, this disordered Kondo singlet state is adiabatically connected
to the usual Kondo spin liquid state \cite{review}.

Comparing the respective ground-state energies $E_g^{KS}$ and $E_g^{AF}$, we
find the phase boundary between the two ground states: 
\begin{eqnarray}
&&\left( \frac 1{\lambda _{\perp }}\right) _c= \coth ^{-1}\left[ \sqrt{%
1+\left( \frac{\lambda _{\parallel }}4\right) ^2}\right.  \nonumber \\
&& \hspace{2cm} \left. +\left( \frac{\lambda _{\parallel }}4\right) ^2\ln
\left( \frac 4{\lambda _{\parallel }}+\sqrt{1+\left( \frac 4{\lambda
_{\parallel }}\right) ^2}\right) \right] ,
\end{eqnarray}
which is displayed by the solid line in Fig.1. Below this line, the AFLRO
state is more stable $E_g^{AF}<E_g^{KS}$; above this line the disordered
Kondo singlet state is more stable. We note that this line intersects the
diagonal $\lambda _{\perp }=\lambda _{\parallel }$ at $0.58$. Taking into
account that the cut-off parameter $D=2t$ with $t$ as the nearest neighbor
hopping, in two dimensions, where our assumption of a constant density of
states is better justified, we find $J_c/t=1.16$. This value is not far from
the numerical results \cite{wang,shi,vekic,assaad} $J_c/t=1.40\sim 1.45$,
obtained for the isotropic model.

Now we turn to discuss the main new result of this paper, namely the
possible coexistence of AFLRO with Kondo singlet state. In fact, our system
of self-consistent equations allows solutions with all order parameters $%
m_d,m_c$ and $V$ being nonzero. Let's start from the region where AF order
dominates, with finite $m_d$ and $m_c$ to see where the instability towards
the Kondo singlet state emerges. Assume $V$ being small, but non-zero, we
obtain the following solution, 
\begin{eqnarray}
&&\left( \frac 1{\lambda _{\perp }}\right) _{c2}=\frac 2{\lambda _{\parallel
}}\frac{m_c}{m_d}  \nonumber \\
&&\text{ \quad }+\sqrt{\frac{1-\frac{m_c}{m_d}}{1+\frac{m_c}{m_d}}}\tan
^{-1}\left( \sqrt{\frac{1-\frac{m_c}{m_d}}{1+\frac{m_c}{m_d}}}\tanh \left( 
\frac{m_c}{m_d}\frac 2{\lambda _{\parallel }}\right) \right) ,
\end{eqnarray}
where $\frac{m_c}{m_d}=\frac{\lambda _{\parallel }}4\ln \left( \frac 4{%
\lambda _{\parallel }}+\sqrt{1+(\frac 4{\lambda _{\parallel }})^2}\right) $.
This instability line for the AF state is delineated by the thick dotted
line in Fig.1. In a similar way, we can also determine the instability
boundary in the Kondo singlet phase, which corresponds to the appearance of
small AF order parameters $m_c$ and $m_d$. From the above self-consistent
equations, we can derive the critical value of the coupling parameter 
\begin{equation}
\left( \frac 1{\lambda _{\parallel }}\right) _{c1}=\frac{\lambda _{\perp
}\sinh \left( \frac 2{\lambda _{\perp }}\right) -2}{4\lambda _{\perp }\left( 
\sqrt{\lambda _{\perp }\sinh \left( \frac 2{\lambda _{\perp }}\right) -1}%
-1\right) },
\end{equation}
and a relation between $m_c$ and $m_d$: $\frac{m_c}{m_d}=\frac{\lambda
_{\parallel }}{4\lambda _{\perp }-\lambda _{\parallel }}$. This instability
line for the disordered Kondo singlet state is displayed by a thin dotted
line in Fig. 1.

Between these two instability lines $(\lambda _{\perp \text{ }c1}<\lambda
_{\perp }<\lambda _{\perp \text{ }c2})$, there is a narrow regime where the
AF and the local Kondo singlet screening order parameters coexist to balance
the energy gain from the transverse and longitudinal exchange couplings. The
disordered Kondo singlet state and the AFLRO state can both be used to
describe two distinct ground states of the conventional Kondo insulators. It
seems to us that a new phase may be present when the exchange coupling
parameters are tuned carefully. In such a new phase, the dynamical magnetic
structure factor should have two contributions: a ${\bf q}$-independent
single site slow component, which is typical of the localized Kondo-type
excitations, and a strongly ${\bf q}$-dependent intersite fast component,
reflecting the magnetic interactions. These features can be detected in
inelastic neutron scattering experiments for some Kondo insulating
materials. Moreover, since the localized spins are {\it partially} screened
by the conduction electrons, and the residual localized spins still have
weak AF long-range correlations mediated by the polarization of the
conduction electrons. This novel feature may be related to the small
magnitude of the induced staggered magnetic moments for URu$_2$Si$_2$ and UPt%
$_3$, the so-called heavy fermion micromagnetism\cite{small}.

We realize that our result is based on a mean field calculation, and its
validity should be further checked by more rigorous analytic and numerical
treatments. Eventually, it should be verified by experiments. We also note
that the numerical results for the isotropic model \cite{wang,shi,assaad}
indicate a second order phase transition between the Kondo singlet and AFLRO
states. However, the order of phase transitions is a very sensitive issue,
and the isotropic case may well be a degenerate point of the more general
anisotropic model we consider here.

In conclusion, we have considered a half-filled anisotropic Kondo lattice
model within a mean field theory. The ground-state phase diagram has been
calculated. In addition to the AFLRO phase and the Kondo singlet phase, we
have found that both of these two distinct phases can coexist as a result of
the balance between the Kondo screening effects and the magnetic
interactions, which provides a possible new ground state for the Kondo
insulating compounds.

\smallskip {\bf Acknowledgments}

One of the authors (G. -M. Zhang) wishes to thank Zhi-Liang Cao, Qiang Gu,
and Xiao-Bin Wang for their useful discussions and help, and would also like
to express his gratitude to International Center for Theoretical Physics
(Trieste, Italy) for the hospitality, where this work was initiated.

\smallskip

Figure Caption

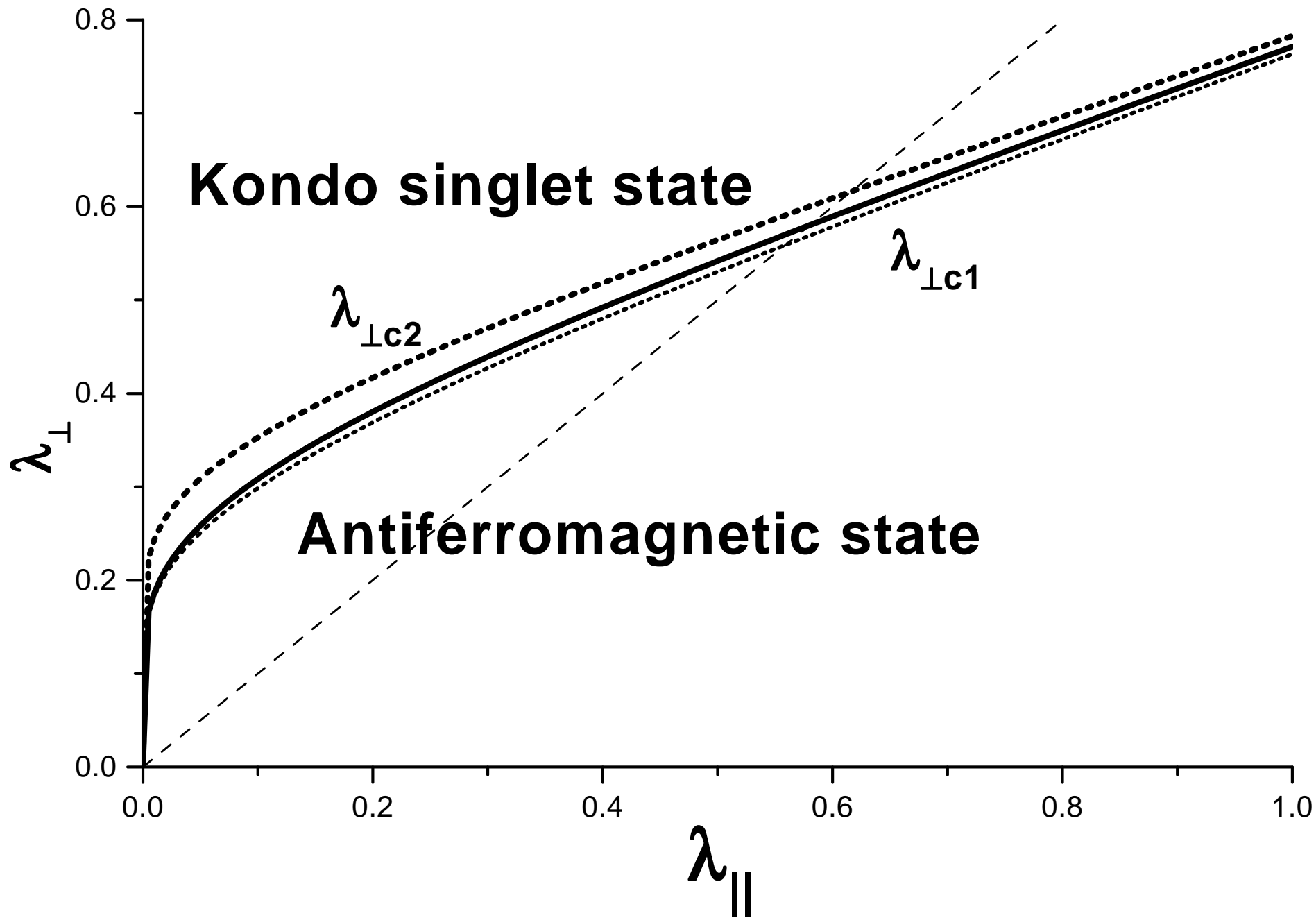
{Fig.1. The ground-state phase diagram of the half-filled anisotropic Kondo
lattice model in the }$\lambda ${$_{\perp }-{\lambda }_{\parallel }$ plane.
The narrow area between the lines of }$(\lambda $$_{\perp })_{c1}$ and $%
(\lambda ${$_{\perp })_{c2}$ is the regime where the local Kondo singlet
state and the AFLRO may coexist. The solid line corresponds to the boundary
of }$E_g^{AF}=E_g^{KS}$, and the thin dashed line denotes the isotropic
limit of the model.

\end{document}